\documentclass[%
 reprint,
 amsmath,amssymb,
 aps,
prab,
]{revtex4-2}

\usepackage{graphicx}
\usepackage{dcolumn}
\usepackage{bm}
\usepackage{hyperref}
\usepackage{amsmath}

\usepackage[authormarkup=none]{changes}

\begin{document}

\preprint{APS/123-QED}

\title{High efficiency uniform positron beam loading in a hollow channel plasma wakefield accelerator}

\author{Shiyu Zhou}
\affiliation{Department of Engineering Physics, Tsinghua University, Beijing 100084, China}
\author{Jianfei Hua}
\affiliation{Department of Engineering Physics, Tsinghua University, Beijing 100084, China}
\author{Wei Lu}
\email[]{weilu@tsinghua.edu.cn}
\affiliation{Department of Engineering Physics, Tsinghua University, Beijing 100084, China}
\author{Qianqian Su}
\affiliation{University of California Los Angeles, Los Angeles, California 90095, USA}
\author{Warren B. Mori}
\affiliation{University of California Los Angeles, Los Angeles, California 90095, USA}
\author{Chan Joshi}
\affiliation{University of California Los Angeles, Los Angeles, California 90095, USA}


\begin{abstract}
  We propose a novel positron beam loading regime in a hollow plasma channel that can efficiently accelerate $e^+$ beam with high gradient and narrow energy spread. In this regime, the $e^+$ beam coincides with the drive $e^-$ beam in time and space and their net current distribution determines the plasma wakefields. By precisely shaping the beam current profile and loading phase according to explicit expressions, three-dimensional Particle-in-Cell (PIC) simulations show that the acceleration for $e^+$ beam of $\sim$nC charge with $\sim$GV/m gradient, $\lesssim$0.5\% induced energy spread and $\sim$50\% energy transfer efficiency can be achieved simultaneously. Besides, only tailoring the current profile of the more tunable $e^-$ beam instead of the $e^+$ beam is enough to obtain such favorable results. A theoretical analysis considering both linear and nonlinear plasma responses in hollow plasma channels is proposed to quantify the beam loading effects. This theory agrees very well with the simulation results and verifies the robustness of this beam loading regime over a wide range of parameters. 
\end{abstract}

\maketitle


\section{Introduction}
\label{Sec:1}

High energy particle physics demands next generation $e^+ e^-$ colliders at the energy frontier for exploring new physics \cite{Michizono2019,Lou2019,Benedikt2019}. To reach the unprecedented beam energy, advanced accelerator concepts that provide a much higher accelerating gradient than the conventional radio-frequency accelerators are especially attractive \cite{Tajima1979,Chen1985,Gai1988}. Among them, plasma wakefield acceleration (PWFA) has achieved great success in $e^-$ acceleration in the so-called blowout regime \cite{Rosenzweig2005,Lu2006PRL,Lu2006POP,Tzoufras2008}. Tens of GeV energy gain has been obtained in a meter-scale plasma\cite{Blumenfeld2007}, and efficient acceleration of an $e^-$ bunch with narrow energy spread has been demonstrated by properly beam loading a trailing $e^-$ bunch behind the drive beam \cite{Litos2014}. However, for $e^+$ bunch acceleration, no beam loading scenario that simultaneously satisfies the requirements of small energy spread of the accelerating positron  beam and high energy extraction efficiency from the drive beam to the trailing beam has been demonstrated.

The performance of a particle collider is mainly quantified by the beam energy and luminosity $L$, that is roughly given by $L({\rm cm}^{-2} {\rm s}^{-1} )=fN^2/4\pi \sigma_x \sigma_y$. Here, $f$, $N$, $\sigma_x$, $\sigma_y$ are the repetition rate of colliding bunches, single bunch population and rms beam sizes at the interaction point in $x$ and $y$ directions respectively. A collider that is built for the precision measurements of the Higgs boson will require center-of-mass $e^-/e^+$ beam energy of about 250GeV and $L\sim 10^{34} {\rm cm}^{-2} {\rm s}^{-1}$, that can be achieved by $10^{10}$ particles per bunch with a collision rate 1kHz and $\sigma_x\sigma_y < 8000{\rm nm}^2$ \cite{Joshi2020}. To tightly final focus the beam usually requires high beam quality, such as the beam energy spread much less than 1\%. Various methods have been proposed to address the plasma wakefield $e^+$ acceleration problem for a compact $e^+ e^-$ collider. A hollow laser \cite{Vieira2014} or electron beam \cite{Jain2015} driver was proposed for $e^+$ acceleration in a uniform plasma, and a finite-width plasma driven by an $e^-$ bunch \cite{Diederichs2019,Diederichs2020} was considered for $e^+$ acceleration and transport. Recently, multi-GeV acceleration of $e^+$ particles has been demonstrated in an $e^+$ self-loaded plasma wakefield \cite{Corde2015a}. However, all these regimes are still far from what is needed for a future collider.

Another promising and extensively explored regime is using a hollow plasma channel where the longitudinal field inside the channel is independent of the transverse dimension and the transverse field vanishes when the highly relativistic beams propagate on-axis \cite{Chiou1995,Chiou1998,Schroeder1999,Schroeder2013,Zhou2021}. Hence both positive and negative charged beams can be transported through the channel simultaneously without being deflected. Previous researches have demonstrated the creation of an extended hollow plasma channel \cite{Gessner2016}. However, most relevant work has been in the linear regime, where the accelerating gradient is moderate and the loaded charge is limited. High gradient acceleration of a high charge $e^+$ beam in a hollow plasma channel, that is desired by the plasma-based particle collider, is still an unsolved problem.

In this work we propose a new positron beam loading regime in a hollow plasma channel that can achieve simultaneously uniform, high-gradient and high-efficiency acceleration of a high-charge positron beam. By loading the $e^+$ bunch on top of a precisely tailored $e^-$ bunch, the excited wakefields inside the hollow plasma channel can be uniformly accelerating for that $e^+$ beam. The energy transfer efficiency from the drive $e^-$ bunch to the witness $e^+$ bunch can be quite high when the two beams are in the same wave bucket. And this regime is valid over a wide range of beam charges and accelerating gradients. These merits are verified by series of quasi three-dimentional (3D) PIC simulations with various parameters. Also, a theoretical model considering the nonlinear blowout effect in a hollow plasma channel is developed to comprehend this regime. This paper is organized as follows. In Sec.\ref{Sec:2}, we examine the validity of the linear theory of the hollow plasma channel when the $e^+$ bunch is loaded behind the drive $e^-$ bunch as in the usual way in a drive-trailing bunch configuration employed in a PWFA. The simulations and theoretical analysis show that the predictions of linear theory will fail as we increase $e^+$ bunch charge to nC when gradient is $\sim$GV/m. 
Then we describe the beam loading scenario when $e^-$ and $e^+$ bunches are spatially overlapped with one another in Sec.\ref{Sec:3}. In this case, one can manipulate current profile of the $e^-$ beam instead of the $e^+$ beam to achieve longitudinally uniform beam loading. Extensive numerical simulations show that this scenario supports high-gradient uniform $e^+$ acceleration when the drive $e^-$ beam is precisely tailored. In Sec.\ref{Sec:4} a theoretical description of this beam loading regime is explored, where the dynamics of the plasma electrons is considered and the blowout effect is quantified. The contributions from the linear and nonlinear responses can be estimated through this model and the estimate agrees well with the simulation. Finally, more discussion and the conclusion are presented in Sec.\ref{Sec:5}. 

\section{Validity of the linear theory in a hollow plasma channel}
\label{Sec:2}

Although hollow plasma channels have been widely studied as a promising way of $e^+$ beam acceleration, most proposals are based on the linear regime that assumes the plasma density profile at the channel wall is unperturbed by the drive and witness beam \cite{Schroeder1999,Gessner2016}. When the accelerated charge and/or gradient increase toward the requirements of linear collider, this critical assumption may no longer hold. Detailed examination of the validity of the linear theory is thus necessary for the performance of hollow plasma channel when it serves as an accelerating unit. 

For an $e^+ e^-$ collider with high beam colliding luminosity and compact size, we seek $e^+$ acceleration with a beam-loaded accelerating gradient $>$1GV/m, charge $\sim$1nC and induced energy spread $<1\%$. To accelerate 1nC $e^+$ bunch, we employ a relativistic $e^-$ driver that contains 2nC charge, and the $e^+$ beam is placed at the accelerating phase behind the drive beam as usual. The hollow plasma channel is an annular shape plasma with the electron density described as $n_p (r)=n_0 [H(r-r_i)-H(r-r_o)]$, where $H$ is the Heaviside step function, $r_i$ and $r_o$ are the plasma channel inner and outer radii respectively. The plasma wakefields according to the linear theory \cite{Gessner2016} are the convolution of wake function and beam charge distribution. Longitudinal wakefield $E_z (\xi)$ can be described as 
\begin{equation}
  E_z(\xi)=\int_{-\infty}^{\xi}\rho(\xi')W_z(\xi-\xi')d\xi'.
  \label{Eq:1}
\end{equation}
Here, $\xi\equiv ct-z$, $\rho(\xi)$ is the longitudinal charge density distribution and $W_z(\xi)=\frac{\mathcal{G}k_p^2}{\pi\epsilon_0}\cos(\chi k_p\xi)$ is the zero-order longitudinal wake function. In the expression of wake function, $k_p = \sqrt{\frac{n_p e^2}{\epsilon_0 m_e c^2}}$ is the plasma wavenumber, $\mathcal{G}$ and $\chi$ are geometric quantities, with $\mathcal{G}=\frac{B_{00}}{k_p r_i(2B_{10}+k_p r_i B_{00})}$, $\chi=\sqrt{\frac{2B_{10}}{2B_{10}+k_p r_i B_{00}}}$and $B_{ij} \equiv I_i (k_p r_i ) K_j (k_p r_o )+(-1)^{i-j+1} I_j (k_p r_o ) K_i (k_p r_i)$, where $I_n$ and $K_n$ are modified Bessel functions. Meanwhile, the transverse wakefield $W_{\perp}=E_r-cB_{\phi}$ is almost 0 inside the channel, allowing for the transport of both $e^-$ and $e^+$ bunches.

\begin{figure}[hb]
  \includegraphics[width=8.6cm]{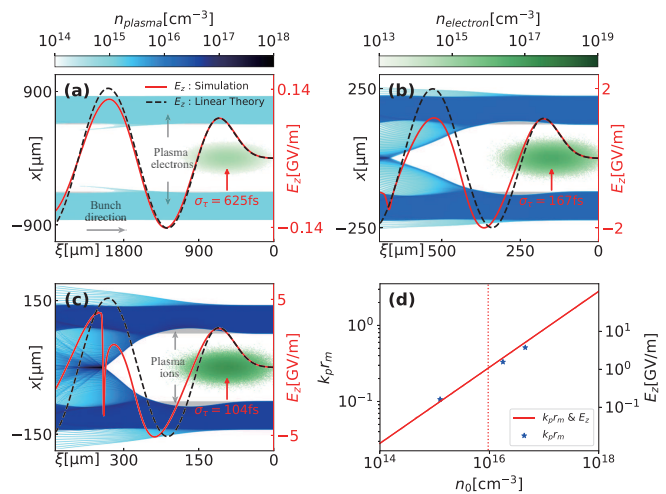}
  \caption{(a-c) Plasma wakes in hollow plasma channels driven by 2nC $e^-$ beams. From (a) to (c) the rms bunch durations are 625fs, 167fs and 104fs and plasma densities are $1.26\times 10^{15}$, $1.77\times 10^{16}$, $4.52\times 10^{16} \rm cm^{-3}$ correspondingly so that the normalized bunch length, size and plasma structures are constant. (d) The maximum normalized outward displacement of the plasma electron sheath and the corresponding maximum $E_z$ against plasma density. The blue asterisks denote $k_p r_m$ for simulations in (a-c). The red dotted line separates the linear and nonlinear regions where the peak $E_z$ field from linear theory is 50\% larger than that in simulation.}
  \label{Fig:1}
\end{figure}

To obtain a larger amplitude plasma wakefield with the given $Q_d=$2nC $e^-$ bunch, a staightforward method is to compress the beam to achieve a higher beam current. Here we select the rms beam length of 625fs, 167fs and 104fs separately, corresponding to the peak current 1.3kA, 4.8kA and 7.6kA respectively. The plasma density is adjusted with the beam length in order that the plasma skin depth is compatible with the beam size. The corresponding plasma response and the wakefields (both in simulation and according to the linear theory) are presented in FIG.\ref{Fig:1}. Simulations in this work are conducted using the quasi-3D PIC code \emph{QuickPIC} \cite{Huang2006,An2013}, and there are $1024\times 1024\times 1024$ cells in $(x,y,\xi)$ dimensions and $6\times 6$ plasma particles in each cell. The plasma densities for the 3 cases in FIG.\ref{Fig:1} are $1.26\times 10^{15}$, $1.77\times 10^{16}$ and $4.52\times 10^{16} \rm cm^{-3}$ so that the normalized rms bunch length $k_p \sigma_z=1.25$ is fixed. So are the normalized bunch size $k_p \sigma_r=0.5$, normalized annular plasma radii  $k_p r_i=3$ and $k_p r_o=5.5$. On-axis $E_z$ lineout is illustrated by the black dashed lines for the linear theory and red lines for the simulation results.

As can be seen in FIG.\ref{Fig:1}, generally higher accelerating gradient wakes are driven in higher density channels by higher current beams, but the simulation results deviate further from the linear theory. In FIG.\ref{Fig:1}a, a tenuous beam only slightly perturbs the wide hollow plasma channel, and drives a moderate amplitude wakefield (peak accelerating gradient $\sim$150MV/m) that agrees well with the linear theory. As we increase the beam peak current to 4.8kA, the plasma electrons near the inner boundary are expelled, then they are pulled back by the more massive and therefore relatively immobile plasma ions and are injected inside the channel as shown in FIG.\ref{Fig:1}b. The peak accelerating field now is only half of the linear theory with the gradient up to 1GV/m. When the 2nC driver is compressed to 104fs as in FIG.\ref{Fig:1}c, the returned plasma electrons gather around the axis and form a density peak. The on-axis $E_z$ is no longer sinusoidal-like, but there is a deep trough during the $e^+$ accelerating phase, which cannot be predicted by the linear theory.

Next, we will elaborate the relationship between the magnitude of accelerating gradient and the validity of the linear theory. If the displacement of plasma electrons is comparable to the plasma skin depth, we would expect that the linear theory no longer holds. For a short drive beam ($k_p \sigma_z \lesssim 1$), the displacement of the plasma sheath electrons is roughly twice the equilibrium blowout radius $r_{eq} (\xi)$ \cite{Lu2005POP}, i.e., the charge of the drive beam is compensated by the ions inside the blowout boundary $n_0 \pi [(r_{eq} (\xi)+r_i )^2-r_i^2 ] = I(\xi)/ec$. So, the normalized maximum blowout radius can be estimated as
\begin{equation}
  k_p r_m \approx \frac{2Q_d k_p}{(2\pi)^{3/2} e r_i \sigma_z n_0}.
  \label{Eq:displacement}
\end{equation}
On the other hand, a drive bunch with a Gaussian profile excites the oscillating $E_z (\xi)$ with the amplitude 
\begin{equation}
  E_{zm}=Q_d \frac{\mathcal{G} k_p^2}{\pi \epsilon_0} e^{-\frac{1}{2} \chi^2 k_p^2 \sigma_z^2}
  \label{Eq:Ezm}
\end{equation}
according to Eq.\eqref{Eq:1}. FIG.\ref{Fig:1}d plots the Eq.\eqref{Eq:displacement} and \eqref{Eq:Ezm} with above parameters. With the fixed normalized quantities, $k_p r_m$ and $E_z$ follow the same trend with respect to the plasma density. The simulated $k_p r_m$ denoted by the asterisks shows a good agreement with Eq.\eqref{Eq:displacement}, that is 0.1, 0.3, 0.5 for the 3 drive beams. As shown in FIG.\ref{Fig:1} (a-c), if $k_p r_m \ll 1$ the linear theory is a reasonable description, but when $k_p r_m \sim 1$ the nonlinear effect plays a major role. As seen in FIG.\ref{Fig:1}d by the dotted red line, when the accelerating gradient $\ge$1GV/m the peak accelerating fields from linear theory and PIC simulation differ by 50\%, which indicates that nonlinear effects should be taken into account.

\begin{figure}[ht]
  \includegraphics[width=8.6cm]{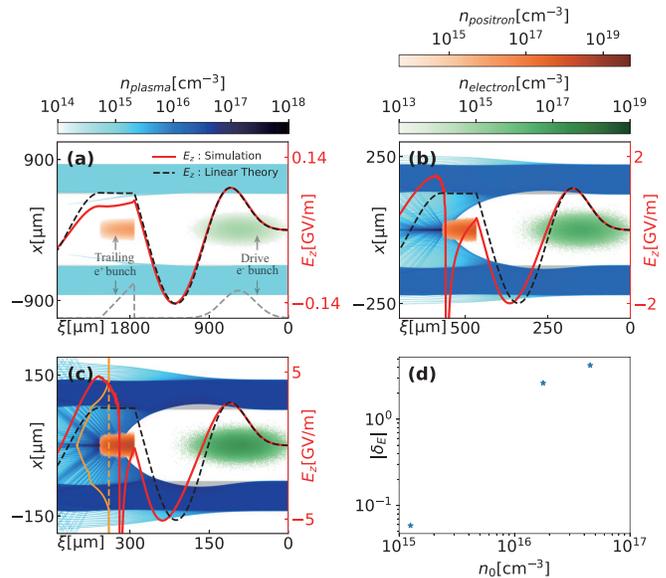}
  \caption{(a-c) Beam loading effects when $e^+$ beam is loaded behind the drive beam. The plasmas and electron beams are identical to that in FIG.\ref{Fig:1}. And the $e^+$ beam profile is chosen to obtain uniform acceleration according to the linear theory. The gray dashed line in (a) is the current profile of the beams. The orange line in (c) is the transverse lineout of $E_z$. (d) The induced energy spread of the $e^+$ beam for these situations.}
  \label{Fig:2}
\end{figure}

The validity of the linear theory with $e^+$ beam loading behind the drive beam must be examined using PIC simulations and is presented in FIG.\ref{Fig:2}. The 1nC $e^+$ beam current and loading phase are chosen so as to flatten the $E_z$ field according to the linear theory as illustrated by the dashed lines in FIG.\ref{Fig:2} and the drive beam and plasma are identical to that in FIG.\ref{Fig:1}. Intuitively, the $e^+$ beam loading suffers more from the nonlinear effects. For the 625fs driver (FIG.\ref{Fig:2}a), the trailing positron bunch doesn't perturb (beam-load) the plasma structure too much, while the loaded $E_z$ is not as uniform as the linear theory prediction. The induced energy spread $\delta_E=\sigma_E/\Delta E$ (the ratio of rms energy to average energy gain) is 5.9\% as plotted in FIG.\ref{Fig:2}d. As the beam load current is increased as in FIG.\ref{Fig:2}b and \ref{Fig:2}c, the plasma electrons are sucked in by the positrons toward the axis, leading to a sharp decreasing field within the positron bunch. $\delta_E$ in these cases could be larger than 100\%. Besides, $E_z$ field is no longer uniform in the transverse direction due to the plasma electrons inside the channel as shown in FIG.\ref{Fig:2}c, that induces extra slice energy spread of the witness beam, which also deviates from the linear theory.

To summarize this section, as we approach the parameter requirements of an $e^+ e^-$ collider, accelerating an $e^+$ beam behind an $e^-$ bunch inside the hollow plasma channel will lose the advantages predicted by the linear theory. However, an important hint from FIG.\ref{Fig:1} and FIG.\ref{Fig:2} is that at the $e^+$ accelerating phase (positive $E_z$) which is the same as the decelerating phase for the drive electrons $E_z$ in simulation agrees much better with the linear theory, which would lead to alternative solutions.

\section{High gradient uniform positron acceleration in a hollow plasma channel}
\label{Sec:3}

 We now examine in detail the loading scenario where the $e^+$ beam coincides with the drive $e^-$ beam, which shows great potential of high gradient uniform positron acceleration. 

In this loading regime, the $e^+$ bunch influences the wakefields by compensating part of the negative charge of the drive beam, and the net current profile now determines the plasma wakefields. To obtain uniform acceleration for the $e^+$ bunch, we can turn to the linear theory of hollow plasma channel as the first step. A systematic way to calculate the current profile for a desired electric field is using the Laplace transformation \cite{Bane1985}. But simple calculus is able to determine the current distribution that maintains uniform longitudinal field within some region. Consider a 2-parts current profile where the front excites the wakefield and the tail part maintains it at a constant level. For demonstration purposes, let the beam current of the 1st part to be uniform $I(\xi)=I_1 (0\le \xi \le L_1)$, and it excites wakefield at $\xi \ge L_1$ according to Eq. \eqref{Eq:1}, $E_{z1}(\xi)=\frac{I_1W_0}{ck_0}\left[\sin(k_0\xi)-\sin k_0(\xi-L_1)\right]$. Here $W_0=\frac{\mathcal{G} k_p^2}{\pi \epsilon_o}$ and $k_0 = \chi k_p$. To maintain a uniform wakefield at $\xi \ge L_1$ so that $E_{z}=\frac{I_1W_0}{ck_0}\sin(k_0L_1)$, the required current profile is linearly ramped $I(\xi)=\frac{I_2}{L_2} (\xi-L_1)+I_3  (L_1\le\xi\le L_1+L_2 )$, where $I_2 = I_1 k_0L_2\sin(k_0L_1)$ and $I_3 = I_1[1-\cos(k_0 L_1)]$. However, a discontinuity at $\xi = L_1$ makes this current profile hard to generate in practice. A linearly ramped current section can be brought in to bridge the gap between these two parts. For a continuous current distribution with the form:
\begin{equation}
  I(\xi)=\left\{
    \begin{array}{cl}
    I_1 & ,\xi \in[0,L_1]
    \\
    \frac{I_2}{L_2}(\xi-L_1)+I_3 & ,\xi \in(L_1,L_1+L_2]
    \\
    \frac{I_4}{L_3}(\xi-L_1-L_2)+I_5 & ,\xi \in(L_1+L_2,\\ & \qquad L_1+L_2+L_3]
    \end{array},
  \right.
  \label{Eq:2}
\end{equation}
the continuity condition demands $I_3 = I_1$,$I_5 = I_1 + I_2$. Let the wakefield in $(L_1+L_2,L_1+L_2+L_3]$ to be constant, then we can obtain the following relations $\frac{I_2}{k_0L_2}\sin(\chi k_pL_2)=-I_1\cos k_0(L_1+L_2)$, $I_1\sin k_0(L_1+L_2)-\frac{I_2}{k_0L_2}(\cos(k_0L_2)-1)=\frac{I_4}{k_0L_3}$ and the uniform wakefield $E_z=\frac{I_2W_0}{ck_0(k_0L_2)}[1-\cos k_0(L_1+L_2)]+\frac{I_3W_0}{ck_0}\sin(k_0L_2)$.

\begin{figure}[ht]
  \includegraphics[width=8.6cm]{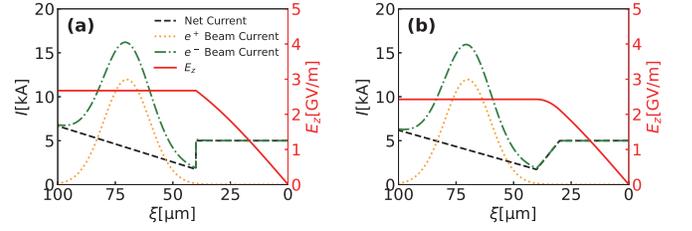}
  \caption{Positron beam loading scenario when the $e^+$ and $e^-$ beams coincide. (a) A discontinuous net current case. (b) A continuous net current case. $E_z$ is calculated according to the linear theory.}
  \label{Fig:3}
\end{figure}

The instances of the above current distributions in a specific hollow plasma channel are illustrated in FIG.\ref{Fig:3}. The wall electron density of the hollow plasma channel is $n_0=1.77\times 10^{16} \rm{cm}^{-3}$, and the inner and outer radii are $45 \rm\mu m$ and $90 \rm\mu m$ respectively. For the 2-part case, $L_1=40 \rm\mu m$, $L_2=60 \rm\mu m$, $I_1=5 \rm kA$, and for a continuous net current profile, $L_1=30 \rm\mu m$, $L_2=10 \rm\mu m$, $L_3=60 \rm\mu m$, $I_1=5 \rm kA$, where other parameters can be obtained through above expressions. The $e^+$ beam has a Gaussian profile in longitudinal direction with the rms length $\sigma_z=10 \rm \mu m$, beam center at $\xi=70 \rm\mu m$ and contains charge of 1nC. The $e^-$ beam current profile is the summation of net current and $e^+$ beam current profile. The on-axis $E_z$ fields presented in FIG.\ref{Fig:3} are calculated according to the linear theory. For both profiles, along the longitudinal direction $E_z$ is increasing at the front $40 \rm\mu m$, then it keeps constant for the rest. The uniform accelerating gradients are 2.67GV/m and 2.42GV/m separately. For the $e^+$ beam, the majority part (within $\pm 3\sigma_z$) resides in the uniform field region. 

\begin{figure*}[ht]
  \includegraphics[width=13cm]{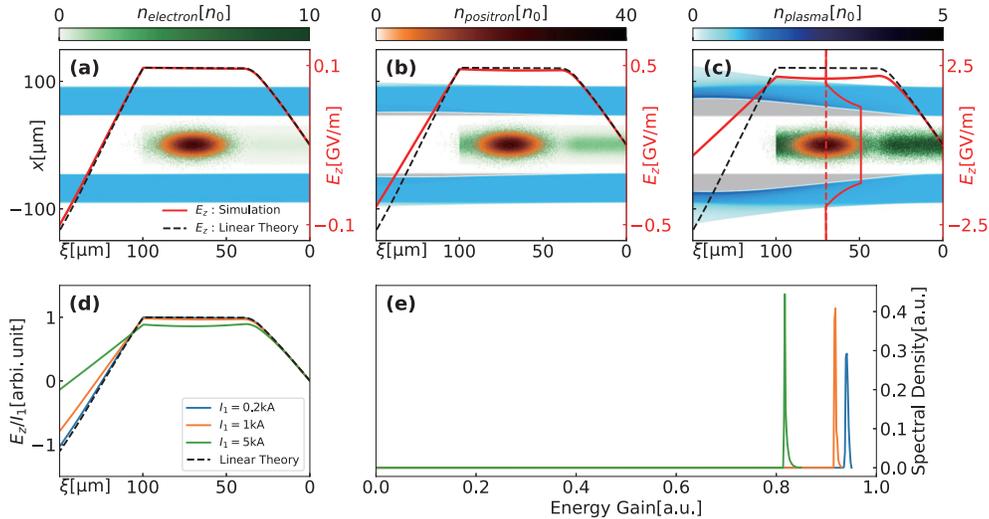}
  \caption{The positron beam loading effects when two bunches coincide. (a-c) The plasma response and the corresponding wakefields for (a) $I_0= \rm 0.2kA$, $Q_{net} \rm =56pC$, (b) $I_0 = \rm 1kA$,$Q_{net} = \rm 280pC$ and (c) $I_0 = \rm 5kA$, $Q_{net}= \rm 1.4nC$. (d) The comparison of $E_z/I_0$ for 3 simulations and the linear theory. (e) The spectrum of the $e^+$ beam energy gain for 3 different situations.}
  \label{Fig:4}
\end{figure*}

The performance of this beam loading regime in practice is evaluated by PIC simulations which can resolve kinetic and nonlinear processes. The intensity of the net current spans a wide range so that it can be examined from linear to nonlinear scenarios. FIG.\ref{Fig:4} illustrates the simulation results. The net current profile is according to Eq.\eqref{Eq:2}, with $I_1=0.2 \rm kA,1kA,5kA$ respectively. $L_1,L_2,L_3$ and the parameters of the 1nC positron beam are identical to that in FIG.\ref{Fig:3}b. In FIG.\ref{Fig:4}a, the linear regime dominates and the simulated $E_z$ almost overlaps with the linear theory prediction. As the net current is weak, the peak accelerating gradient is only 100MV/m. In FIG.\ref{Fig:4}b, the intensity of the net current is increased by 5 times. The uniform accelerating field is now about 500MV/m, and the simulation result also agrees well with the linear theory. Further increase the net current with $I_1=5\rm kA$ as in FIG.\ref{Fig:4}c. It’s obvious that the plasma electrons are strongly blown out by the intense drive beams, and the structure of the hollow plasma channel deviates a lot from the unperturbed situation. Thus, $E_z$ differs between the simulation and linear theory. Surprisingly, at the $e^+$ acceleration region $E_z$ is still flat, although the amplitude is not the same as the linear theory predicts. The $e^+$ beam now gets uniform acceleration of the gradient 2.1GV/m. Besides, since the plasma electrons move outward at the $e^+$ acceleration phase, $E_z$ keeps uniform in the transverse directions inside the channel as plotted in FIG.\ref{Fig:4}c, which leads to a narrow slice energy spread.

The comparisons between the linear theory and simulation results are presented in FIG.\ref{Fig:4}d, where the $E_z$ are normalized according to their net current strength. As the net current increases, the linear theory is more inaccurate to describe the wakefields as expected but the characteristic of uniform $e^+$ beam acceleration maintains. FIG.\ref{Fig:4}e plots the spectrum of the energy gain of the $e^+$ beam, where the induced rms energy spread can be measured as 0.29\%, 0.42\% and 0.55\% for $I_1=0.2 \rm kA,1kA,5kA$ respectively. Since $E_z$ is uniform in the transverse dimension, the induced slice energy spread for all cases is less than 0.1\%, allowing the next step manipulation of the phase space such as energy dechirper to further reduce the total energy spread \cite{Wu2019,Wu2019PRAppl}. Energy transfer efficiency, i.e., the ratio between energy loss of the electron beam and energy gain of the positron beam, for the 3 cases is 95.7\%, 81.5\% and 46.3\% for $I_1=0.2 \rm kA,1kA,5kA$ respectively. In principle, by adjusting the ratio of $e^-$ and $e^+$ beam charges, the energy efficiency can be close to 100\% for any accelerating gradient. 

\begin{figure}[hb]
  \includegraphics[width=8.6cm]{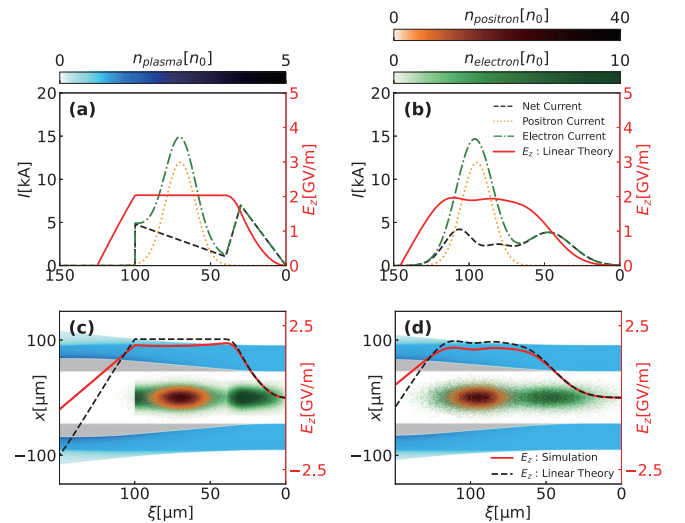}
  \caption{Two other positron beam loading cases. The drive beam has (a) a piece-wise linearly ramped and (b) a bi-Gaussian current profile. The $E_z$ fields are calculated according to the linear theory. (c, d) Simulated plasma responses and wakefields.}
  \label{Fig:5}
\end{figure}

The above analyses and simulations have validated the effectiveness of this fresh beam loading regime for high gradient uniform $e^+$ acceleration. In fact, to achieve the same goal the choice of the current profiles is infinite. FIG.\ref{Fig:5} presents two other instances with more practical current profiles of the $e^-$ beam. FIG.\ref{Fig:5}a illustrates a linearly ramped current profile instead of flat-top at the beam front and the rest part is adjusted according to the linear theory so that the $e^+$ beam gets uniform acceleration. FIG.\ref{Fig:5}b shows a case where the drive beam is with a bi-Gaussian current profile. The $e^+$ beam is still identical to the above cases, and the specific profile of the 2nC drive beam is chosen to optimize the energy spread of the $e^+$ beam. PIC simulations with the proposed beam parameters are conducted, and the results are presented in FIGs.\ref{Fig:5}c, d. As we expected, the main features are consistent with that in FIG.\ref{Fig:4}c. The plasma electrons are expelled from the channel boundary, leading the simulated wakefield to differ from the linear theory, while it remains nearly uniform for $e^+$ beam acceleration. From PIC simulations, we find that the average accelerating gradient for the $e^+$ beam is 1.8GV/m and 1.7GV/m respectively for the two cases, and the induced energy spread is 0.80\% and 1.53\%. The induced slice energy spread for both situations is less than 0.1\%. It is also obvious that fine-tuning of the current profiles can further deduce the correlated energy spread (chirp) to a great extent.

\section{Explanation of the nonlinear beam loading effect}
\label{Sec:4}

Through discussions in Sec.\ref{Sec:3} we have found that the novel positron beam loading regime in hollow plasma channel can achieve high-gradient high-efficiency uniform acceleration. Meanwhile, as mentioned above the plasma structure has non-negligible deformation when the gradient is up to GV/m and nonlinear effects take part in. It is thus important to figure out how the nonlinear phenomena alter the beam loading effect. 

\begin{figure*}[ht]
  \includegraphics[width=13cm]{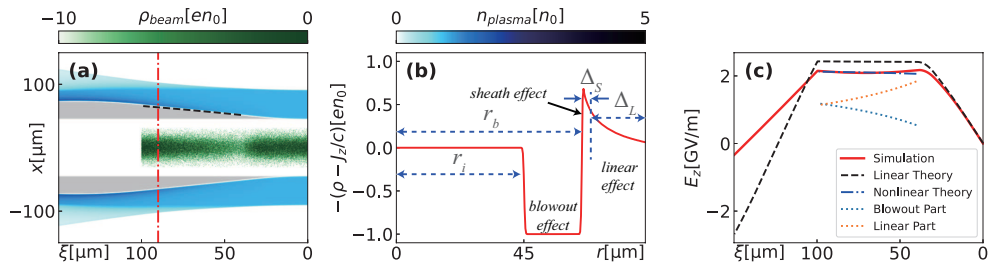}
  \caption{Theoretical analysis of the beam loading regime in a hollow plasma channel. (a) The plasma density and net charge density of the beams with identical parameters to FIG.\ref{Fig:4}c. The black dashed line is the blowout radius obtained from adiabatic approximation. (b) The radial distribution of $-(\rho - J_z/c)$ at $\xi = 90\mu m$ denoted by red line in (a). (c) The on-axis $E_z$ lineouts from the simulation, linear and nonlinear theory.}
  \label{Fig:6}
\end{figure*}

To fully resolve the plasma wakefields the kinetics of the plasma electrons should be carefully considered. In plasma based acceleration, the wakefields can be determined by the pseudo or wake potential $\psi$, which depends on the distribution of the plasma and beams\cite{Lu2006PRL}. $\psi$ is defined as $\psi(r,\xi) \equiv \phi-cA_z$, where $\phi$ and $A_z$ are the scalar and vector potential of the electromagnetic fields. In PWFA, $\psi$ obeys 2D Poisson equation $\nabla_\perp^2 \psi = -\frac{1}{\epsilon_0} (\rho-J_z/c)$, that $\rho$ and $J_z$ are the charge density and current density along longitudinal direction for all charged particles. For the highly relativistic beams propagation along $z$, $\rho-J_z/c=\rho(1-v_z/c)\approx 0$, so the wake potential is determined by the plasma.

For the same physical and simulation parameters as in FIG.\ref{Fig:4}c, the plasma and beam profiles are presented in FIG.\ref{Fig:6}a. The distribution of $S \equiv -(\rho -J_z/c)$ obtained from PIC simulations at $\xi = 90\rm \mu m$ as indicated by the red dash-dot line is illustrated in FIG.\ref{Fig:6}b. At this slice, the plasma electrons in the wall of the channel have been blown out by the drive beam, leaving behind the massive plasma ions and forming a narrow sheath of the electrons. Thus $S$ can be decomposed into several parts according to these physical features. Inside the channel, there are no plasma particles and $S$ is equal to 0. Within the blowout radius $r_b$ of the plasma electrons, $S$ is determined by the exposed plasma ions $S=-en_0$. For $r>r_b$, there is a narrow sheath of plasma electrons with high density. Further away, it is mainly the linear response of the plasma electrons that are accelerated by the plasma wakefield and contribute to the plasma current. The profile of $S$ is very similar to that in the blowout regime in a uniform plasma except for the hollow region\cite{Lu2006PRL}. Correspondingly, the $\psi$ can also be identified by these separate contributions

\begin{widetext}
\begin{eqnarray}
  \label{Eq:3}
  \psi (0, \xi) &=& \frac{1}{\epsilon_0} \int_0^\infty \frac{dr}{r} \int_0^r (\rho - J_z/c)r' dr'  \nonumber \\
  &=& \frac{1}{\epsilon_0}  \left\{ \int_0^{r_b} + \int_{r_b}^{r_b + \Delta_s} + \int_{r_b + \Delta_s}^\infty \right\}  \frac{dr}{r} \int_0^r (\rho - J_z/c)r' dr' \nonumber    \\
  &=& \psi_b + \psi_s + \psi_{linear}.
\end{eqnarray}
\end{widetext}

With the distribution of $S$, the pseudo-potential can be obtained by integration of $S$ as in Eq. \eqref{Eq:3}. Since $S$ has distinct forms for different regions, $\psi$ can also be identified by these separate effects, i.e., the blowout effect, sheath effect and linear effect. A critical parameter in Eq. \eqref{Eq:3} is the blowout radius $r_b$ of the plasma electrons. At the uniform acceleration region, the net current $I(\xi)$ slowly increases along $\xi$, and so does the blowout radius. Assume the evolution of those plasma electrons in the electron sheath is adiabatic, then the blowout radius is the equilibrium radius $\pi(r_b^2 (\xi)-r_i^2 )en_0=I(\xi)/c$. The estimated $r_b (\xi)$ is plotted in FIG.\ref{Fig:6}a by the black dashed line, that shows a good agreement with the simulation. With the expression of  $r_b (\xi)$, $\psi_b$ can be easily calculated 
\begin{equation}
  \begin{aligned}
    \psi_b (0,\xi) &= \frac{1}{\epsilon_0} \int_{r_i}^{r_b} \frac{dr}{r} \int_{r_i}^r en_0 r' dr' \\
    &=  \frac{e n_0}{\epsilon_0}  \left[ \frac{1}{4} (r_b^2 (\xi)-r_i^2) - \frac{1}{2} r_i^2 \ln \frac{r_b (\xi)}{r_i}\right]
  \end{aligned}
  \label{Eq:psi_b}
\end{equation}
for $\xi \in [40,100]\mu m$. $E_z$ is given by $E_z  \approx \frac{\partial}{\partial\xi} (\phi - cA_z )= \frac{\partial \psi}{\partial\xi} = \frac{\partial \psi_b}{\partial\xi} + \frac{\partial \psi_s}{\partial\xi}+\frac{\partial \psi_{linear}}{\partial\xi}$. So, the contribution to $E_z$ by the blowout effect is 
\begin{equation}
  \begin{aligned}
    E_{zb}(\xi) &=\frac{d\psi_b(\xi)}{d\xi} = \frac{1}{4} \frac{en_0}{\epsilon_0} \frac{dr_b^2 (\xi)}{d\xi} \left(1- \frac{r_i^2}{r_b^2(\xi)}\right)\\
    &= \frac{1}{4\pi c\epsilon_0} \frac{dI(\xi)}{d\xi} \left(1- \frac{r_i^2}{r_b^2(\xi)}\right),
  \end{aligned}
  \label{Eq:Ez_b}
\end{equation}
which can be explicitly calculated as Eq.\eqref{Eq:Ez_b} and is presented in FIG.\ref{Fig:6}c. As in the blowout regime of a uniform plasma, the plasma sheath is narrow and $\psi_s$ is much less than $\psi_b$ \cite{Lu2006POP}. Thus, the sheath effect is neglected here. As for the contribution of the plasma linear response, it can be approximated by modifying the linear theory. In the linear theory of hollow plasma channel, the geometrical quantity $\mathcal{G}$ is related to the inner radius of the plasma $\mathcal{G} = \frac{B_{00}}{k_p r_i (2B_{10}+k_p r_i B_{00})} \propto \frac{1}{r_i^2} $. For the expected uniform acceleration region, we estimate the linear contribution by 
\begin{equation}
  E_{zl} \approx E_{zl0}  \frac{r_i^2}{(r_b (\xi)+ \Delta_s (\xi))^2},
  \label{Eq:Ez_l}
\end{equation}
where $E_{zl0}$ is the uniform gradient obtained from the linear theory for an unperturbed hollow plasma channel. The on-axis $E_z (\xi)$ can be expressed by $E_z = E_{zb} + E_{zl}$.

If we assume $\Delta_s = 0.1 r_b$, $E_z (\xi)$ can then be explicitly quantified and the result is presented in FIG.\ref{Fig:6}c by the blue dash-dotted line. That field is rather flat at the $e^+$ beam acceleration region and the amplitude matches the simulation result as well. The contributions from the blowout effect and plasma linear response are also presented, which gives a qualitative explanation of the uniform acceleration. As the plasma electrons are blown out by the drive beams, the inner radius of the hollow plasma channel is effectively wider, leading to the reduction of the plasma linear wakefields. At the same time, the blowout of the plasma electrons contributes to the longitudinal wakefield and partially compensates the reduction. As a result, at the $e^+$ beam acceleration region, the accelerating gradient is smaller than the linear theory, while it is still rather flat. Also note that when $I\ll \pi a^2 en_p c$, $r_b \approx r_i$ and $\psi \approx \psi_{linear}$, i.e., the contribution of the nonlinear response is negligible and linear theory is enough to describe the wakefield structure.

\section{Discussion and conclusion}
\label{Sec:5}

Uniform high-gradient acceleration of a high-charge positron beam is critical for the next generation plasma based $e^+ e^-$ collider. In this work, we propose a novel beam loading scenario in the hollow plasma channel to address this problem. By combining the high charge $e^-$ and $e^+$ beams with specific current profiles, the $e^+$ beam can obtain high-gradient acceleration with narrow energy spread and high energy transfer efficiency. Series of PIC simulations show that the acceleration for $e^+$ beam of $\sim$nC charge with $\sim$GV/m gradient, $\sim$0.5\% induced energy spread and $\sim$50\% energy transfer efficiency is possible. Fine-tuning of the current profiles can further decrease the projected energy spread. Besides, there is no constraint on the $e^+$ beam current distribution, as long as the $e^-$ beam current profile is precisely tailored, which is more practical in experiments \cite{Lemery2014,Loisch2018}.

 For higher energy gain of the $e^+$ beam, transportation over a long distance is necessary where the guiding of the $e^+$ and/or $e^-$ beam and the stability could be important issues \cite{Lindstrom2018}. External quadrupole magnets can be employed to guide both the drive and witness beams as proposed by previous researches \cite{Caldwell2009,Yi2014}. And the emittance preservation will set stringent limits on beam misalignment with respect to the hollow plasma channel.

As the $e^+$ beam coincides with the $e^-$ beam, there is a chance a positron annihilates with an electron, and the rate is related to their beam density and center-of-mass energy. In general, the rate of electron-positron annihilation can be estimated as $\Gamma \lesssim \pi r_0^2 c n_e$, where $r_0$ is the classical radius of the electron and $n_e$ is the electron density\cite{Greaves2002}. Typically, in PWFA the electron beam density is on the order of $10^{17} \rm cm^{-3}$ or less. With $n_e=10^{17} \rm cm^{-3}$, the annihilation time $\tau_a = 1/ \Gamma \sim 1\rm ms$, during which time the propagation distance is about hundreds of kilometers. Thus, for the acceleration distance less than 1 kilometer, the annihilation due to the drive $e^-$ beam is negligible.

The nonlinear theory discussed above is not a fully self-consistent nonlinear theory such as the theory of blowout regime\cite{Lu2006PRL}. But this heuristic analysis also reveals important characteristics for the nonlinear regime in hollow plasma channels, that the plasma linear response is of equal importance to the blowout effect in this regime. While in the blowout regime of uniform plasma, the nonlinear effects usually dominate the plasma wakefields. This work will be helpful for the development of a self-consistent nonlinear theory in a hollow plasma channel.

\bibliography{apssamp}

\end{document}